# Vision-Based Classification of Social Gestures in Videochat Sessions

**Yuan Yao**
Computer Science
University of Minnesota
Minneapolis, MN, USA
yaoxx340@umn.edu

**Svetlana Yarosh**
Computer Science
University of Minnesota
Minneapolis, MN, USA
lana@umn.edu

**ABSTRACT**
This paper describes the design and evaluation of the vision-based classification of social gestures, such as handshake, hug, high-five, etc. This is a component of the mediated social touch systems, which can be incorporated into ShareTable and SqueezeBands system to achieve automated gestures recognition and transmission of the touch between the users in real time. The results from our pilot study show the recognition accuracy of each gestures, and they indicate that significant future work is necessary to improve its practical feasibility in the mediated social touch applications.

## 1. INTRODUCTION

Mediated Social Touch (MST) is a technological paradigm that focuses on allowing people to touch across distance in order to reinforce social relationships [1]. The video-mediated communication like videochat allows users to stay in touch with their family and friends by capturing and transferring of video images and audio sounds. Some researches that focus on enhancing communication experience incorporate a haptic channel into the traditional audio and video channels. Users are able to feel the the sense of touch that is created by their partners initially and simulated by the haptic communication devices that applies forces or motion to the users. Therefore, these devices should have the capability to understand the human gestures. Gesture recognition is a method to interpret human gestures via mathematical algorithms, and we develop a hand gesture recognition system that can be incorporated into ShareTable and Squeezebands. In addition to the standard videoconferencing that allows audio and video connection, the ShareTable has a projector-camera system which superimposes a video stream of one table's surface on top of the other to provide a shared workspace. This system is described in detail in others' publications [2]. SqueezeBands haptic mediated social touch system consists of a two sets of haptic bands for each user to during the videochat. When the one of the user make a particular gesture, the system will be activated and transmit the touch to another user through pressure and heat using Shape Memory Alloy haptics. Since SqueezeBands supports five types of gestures: handshake, high-five, hug, shoulder pat, and holding hands, we build a hand tracking and gesture recognition system to support automated detection of these gestures.

We begin by discussing the implementation of the hand tracking and gesture recognition system. We also describe the settings and procedures of the pilot study. Finally, we report the gesture detection accuracy of this system and discuss some potential ways to improve its feasibility.

## 2. TECHNICAL APPROACH

The initial study of the SquuzeBands used Wizard-of-Oz to achieve gesture recognition, this approach is not scalable in the field. Adding vision based hand gesture recognition to haptic band system is necessary for deployment outside of a controlled setting (particularly moving towards field deployments necessary to account for novelty effects). Towards this goal, we developed a hand tracking and gesture recognition system using the Kinect v2 sensor (Figure 1), which integrates a RGB camera and an infrared laser projector depth sensor. The Kinect SDK provides support for developing Windows applications such as human movement tracking and recognition using skeletal tracking and distance measurement using depth data.

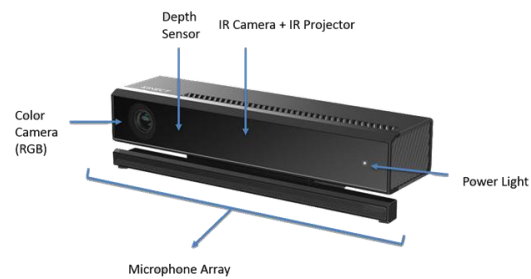

Figure 1. Kinect v2 sensor [4]

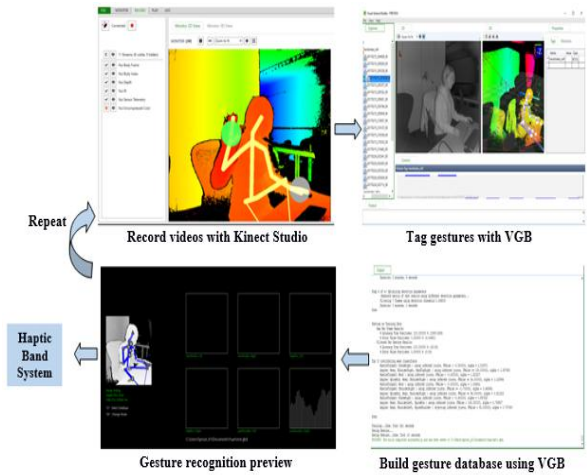

Figure 2: The process of building the database

*2.1 Hand Tracking with Kinect*
We used a freestanding Kinect to implement a hand tracking system to locate users' hands and use these locations to classify gestures. Our C# application, initializes the sensor and access the color, depth, infrared, and body streams from the Kinect. The Kinect allows us to track 25 joints in each body. After getting access to this tracking, we were able to obtain the three-dimensional positions and the rotation information of both hands and thumbs of the participants. The system detects specific hand states of both hands (hand on table, hand extended with palm out, hand extended towards front camera with hand curved, hand reaching toward side of monitor), which allow us to detect each of the five gestures which are handshake, high-five, hug, shoulder pat, and holding hands. The next section describes the classifier we built to automatically recognize each gesture.

*2.2 Gesture Recognition with Visual Gesture Builder*
Visual Gesture Builder (VGB) in Kinect SDK is a tool that provides a solution to gesture recognition through machine learning [3]. To train the recognizer, we recruited 12 undergraduate student volunteers to record Kinect raw chips in which they performed specific gestures. We refer to the eight gestures we trained with the following abbreviations: high-five with right hand (R5), high-five with left hand (L5), handshake with right hand (RH), handshake with left hand (LH), patting shoulder with right hand (RS), patting shoulder with left hand (LS), placing right hand on the mat (RM), and placing left hand on the mat (LM). We did not include the "hug" as a separate gesture since it is simply a combination of the LS and RS gestures done simultaneously. We labeled and tagged all the frames in the recordings that define a gesture, to build a solution and create a database, which could be used with the SqueezeBand and ShareTable system (or other videochat system) at run time. We conducted several iterations of training and testing using Live Preview in VGB, which shows a prediction and prediction confidence of each gesture live. The process of building the database is shown in Figure 2.

**3. EXPERIMENT AND RESULTS**
After we were satisfied with the training data set, we conducted a constrained, lab-based evaluation of our gesture detection classifier.

*3.1 Pilot Study*
Seventeen undergraduate and graduate student volunteers tested this hand tracking and gesture recognition system. Each participant was instructed in each of the gestures and provided a brief demo of the system. When they were ready to start, the computer randomly selected a gesture name and showed them on the screen one at a time. Each participant was instructed to generate 27-30 specific gestures (each lasting 7 seconds).

*3.2 Initial Results*
We compared the gestures that the participants were asked to produce with the gesture that was detected by the classifier. We want to note that we are omitting the RM and LM (hand on mat) from this analysis. While the overall recognition rate for these gestures was fairly high (RM = 88% and LM = 90%), several of the participants placed their hand on the table during the study and kept one hand on the table while doing other gestures, which made the detection of this gesture function differently from the others. The RM and LM gestures are only necessary for the "hold hands" gesture and this gesture is only relevant when combining SqueezeBands with the ShareTable system (versus standard videochat). Given that the ShareTable system currently provides accurate 20-point multi-touch detection for locating a hand on the projection area, the classifier for this particular gesture may be redundant. The recognition rates obtained for the six retained gestures are shown in the form of confusion matrix in Figure 6. The overall recognition rate for these six gestures is 89%. While the accuracy of detecting handshakes with right hand and patting shoulder with left hand are relatively lower than other gestures, the general accuracy of the recognition was fairly high. This follow-up study presents the preliminary feasibility of automatically detecting gestures directed to a videochat partner for the purpose of triggering a haptic device. Three factors may increase or decrease this accuracy in the field. First, gestures enacted naturalistically may have lower accuracy than ones enacted in response to an instruction because there may be contextual factors influencing the specifics of how a user moves. Additionally, in the field the system will have to differentiate not just between the six types of gestures but also whether any gesture was attempted at all. However, there are also two factors that may also increase accuracy. First, we could incorporate a brief user-specific training session when installing the SqueezeBands system. A user-specific classifier is likely to have higher accuracy than a general classifier employed by our system. Second, two-thirds of the gestures (L5, R5, LH, RH) classified are dyadic in nature. For example, a handshake should only be classified as a handshake if the system is reasonably confident that *both* participants are attempting the same gesture. Thus, high confidence for one of the participants may make up for low confidence for the other participant in selecting the correct gesture. Overall, despite the initial promise shown by our preliminary gesture detection system, significant future work is necessary to demonstrate its effectiveness in the field.

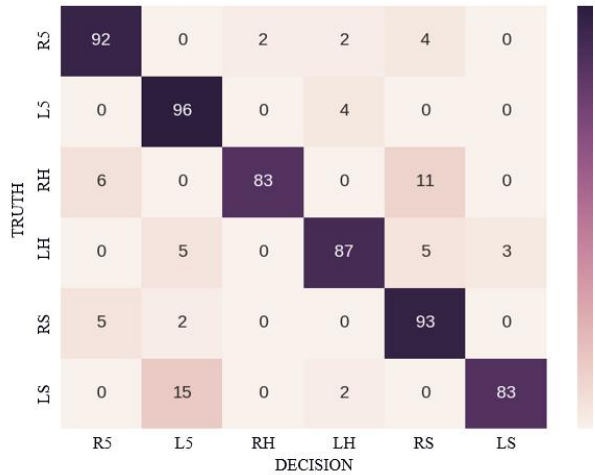

Figure 3. Gesture recognition confusion matrix for the 6 gestures retained (numbers reflect percent of true gestures).

## 4. DISCUSSION

This paper discusses a hand gesture recognition system which can be deployed on the Mediated Social Touch system. The system is developed and tested successfully with Kinect v2 sensor and can be directly incorporated into ShareTable and SqueezeBands. All gestures have recognition rate in between 80-100%, and overall accuracy of this system is approximately 90%. However, the accuracy of some gestures like handshake with right hand and patting shoulder with left hand should be increased before putting this system into use. Some methods we can use in the future include training gestures with more volunteers to create a larger database. Also, in addition training skeleton data using adaBoost, we can use some other features to trains the models using different deep learning algorithms.